\documentclass[11pt]{article}
\usepackage{amsmath,amssymb,epsf}

\def\crampest{\medmuskip = 1mu plus 1mu minus 1mu}

\let\a=\alpha

\def\nn{\nonumber}

\let\bm=\bibitem

\newcommand{\be}{\begin{equation}}
\newcommand{\ee}{\end{equation}}
\def\ba{\begin{array}}
\def\ea{\end{array}}
\def\ft#1#2{{\textstyle{\frac{\scriptstyle #1}{\scriptstyle #2}}}}
\def\fft#1#2{\frac{#1}{#2}}
\def\del{\partial}

\def\sst#1{{\scriptscriptstyle #1}}

\def\td{\tilde}
\def\wtd{\widetilde}
\def\ie{\rm i.e.\ }
\def\dalemb#1#2{{\vbox{\hrule height .#2pt
        \hbox{\vrule width.#2pt height#1pt \kern#1pt
                \vrule width.#2pt}
        \hrule height.#2pt}}}

\newcommand{\hoch}[1]{$\, ^{#1}$}
\newcommand{\bea}{\begin{eqnarray}}
\newcommand{\eea}{\end{eqnarray}}

\def\0{{\sst{(0)}}}
\def\1{{\sst{(1)}}}
\def\2{{\sst{(2)}}}
\def\3{{\sst{(3)}}}
\def\4{{\sst{(4)}}}
\def\5{{\sst{(5)}}}
\def\6{{\sst{(6)}}}
\def\7{{\sst{(7)}}}
\def\8{{\sst{(8)}}}

\def\cA{{{\cal A}}}

\def\im{{{\rm i}}}

\def\R{\rlap{\rm I}\mkern3mu{\rm R}}

\def\R{\rlap{\rm I}\mkern3mu{\rm R}}

\def\R{{{\mathbb R}}}

\def\CP{{{\mathbb C}{\mathbb P}}}
\def\RP{{{\mathbb R}{\mathbb P}}}
\def\Z{{{\mathbb Z}}}

\def\gcd{{\rm gcd}}
\def\kdel#1{{\fft{\del}{\del#1}}}
\begin{document}

\begin{flushright}
MIFP-06-13 \\
{\bf hep-th/0605222}\\
May\  2006
\end{flushright}

\begin{center}

{\Large {\bf Resolutions of Cones over Einstein-Sasaki Spaces}}

\vspace{20pt}

H. L\"u and C.N. Pope

\vspace{20pt}

{\hoch{\dagger}\it George P. \&  Cynthia W. Mitchell Institute
for Fundamental Physics,\\
Texas A\&M University, College Station, TX 77843-4242, USA}

\vspace{40pt}

\underline{ABSTRACT}
\end{center}

     Recently an explicit resolution of the Calabi-Yau cone over the
inhomogeneous five-dimensional Einstein-Sasaki space $Y^{2,1}$ was
obtained.  It was constructed by specialising the parameters in the
BPS limit of recently-discovered Kerr-NUT-AdS metrics in higher
dimensions.  We study the occurrence of such non-singular resolutions
of Calabi-Yau cones in a more general context.  Although no further
six-dimensional examples arise as resolutions of cones over the
$L^{pqr}$ Einstein-Sasaki spaces, we find general classes of
non-singular cohomogeneity-2 resolutions of higher-dimensional
Einstein-Sasaki spaces.  The topologies of the resolved spaces are of
the form of an $\R^2$ bundle over a base manifold that is itself an
$S^2$ bundle over an Einstein-K\"ahler manifold.

{\vfill\leftline{}\vfill \vskip 10pt \footnoterule {\footnotesize
Research supported in part by DOE grant
DE-FG03-95ER40917.
}


\newpage
\tableofcontents
\addtocontents{toc}{\protect\setcounter{tocdepth}{2}}

\section{Introduction}

    The study in string theory of D3-branes located at the apex of a
Calabi-Yau cone has provided a powerful motivation in recent years for
studying complete and non-singular Calabi-Yau metrics that can be
regarded as resolutions of the singular cone metrics.  The essential
idea is to seek a metric which, at large distance, is asymptotic to
the cone metric $ds^2 = dr^2 + r^2 d\bar s^2$ over an Einstein-Sasaki
space $X$ equipped with metric $d\bar s^2$, but which, at small
distance, deforms in such a way that the singular apex $r=0$ of the
cone metric is smoothed out into a non-singular endpoint.

    Typically, this smoothing out of the apex occurs because the level
surfaces at constant radius, which are in general smooth $r$-dependent
deformations of the Einstein-Sasaki space $X$, degenerate at some
small radius $r=r_0$ in a such a way that one obtains a regular
``closing off'' of the manifold at an origin of the radial coordinate
at $r=r_0$.  A classic example is the four-dimensional Eguchi-Hanson
metric \cite{egha}, for which the level surfaces are $\RP^3=S^3/\Z_2$
\cite{begipapo}.  The $\RP^3$ surfaces are ``squashed'' along the
$U(1)$ Hopf fibre over $S^2$ as a function of radius, approaching the
round $\RP^3$ at infinity, and degenerating to $S^2$ at some inner
radius $r_0$ where the Hopf fibre shrinks to zero. This degeneration
is like the origin of polar coordinates on the plane, implying that
the whole Eguchi-Hanson manifold is an $\R^2$ bundle over $S^2$.  It
can be viewed as a resolution of the Calabi-Yau cone over $\RP^3$.

   Countless further examples of this kind are known in the
literature, including, for instance, metrics on 2-plane bundles over
arbitrary Einstein-K\"ahler base spaces $B$ \cite{berber,pagpop}.
They can be viewed as resolved Calabi-Yau cones over Einstein-Sasaki
spaces $X$ that are themselves $U(1)$ bundles over the
Einstein-K\"ahler spaces $B$.  Another important example, in six
dimensions, is the so-called ``deformed conifold,'' which is an $\R^3$
bundle over $S^3$ \cite{conifold}.  Further examples include resolved
cone metrics with $G_2$ or Spin(7) holonomy in seven or eight
dimensions \cite{brysal,gibpagpop}.

   In almost all the explicit examples of resolved cones, the metrics
have cohomogeneity 1, with level surfaces at constant radius $r$ that
are themselves homogeneous spaces.  The ``shape'' of the level
surfaces distorts as a function of $r$, whilst maintaining the
homogeneity.

   In a recent paper, a six-dimensional example of a smooth resolution
of a Calabi-Yau cone was obtained \cite{ootyas} in which the metric
has cohomogeneity 2, meaning that the level surfaces are themselves
inhomogeneous spaces of cohomogeneity 1.  In fact, the example in
\cite{ootyas} was obtained by taking a local class of BPS limits of
six-dimensional Kerr-NUT-AdS metrics that were obtained in a general
construction in \cite{chlupo3}\footnote{The cosmological constant of
the Kerr-NUT-AdS metric scales to zero in the process of taking the
BPS limit.}, and showing that for an appropriate specialisation of the
parameters, a complete and non-singular metric is obtained.  The
smooth metric in \cite{ootyas} can be viewed as a resolution of the
Calabi-Yau cone over the Einstein-Sasaki space $Y^{2,1}$, which is a
special case of the $Y^{p,q}$ class of Einstein-Sasaki spaces
constructed in \cite{gamaspwa}.

   In fact the BPS limits of the even-dimensional Kerr-NUT-AdS metrics
obtained in the general construction in \cite{chlupo3} can all be
viewed locally as deformations of cones over the $(2n+1)$-dimensional
Einstein-Sasaki metrics $L^{p,q,r_1,\ldots,r_{n-1}}$ that were
constructed in \cite{cvlupapo1,cvlupapo2}.  In the special case of six
dimensions, the general BPS construction in \cite{chlupo3} gives
cohomogeneity-3 metrics that are cones over the $L^{pqr}$ class of
five-dimensional Einstein-Sasaki metrics obtained in
\cite{cvlupapo1,cvlupapo2}.

   In view of the above observations, it is natural to investigate the
possibility of generalising the result in \cite{ootyas}, to see
whether the classes of BPS limits of Kerr-NUT-AdS metrics considered
in \cite{chlupo3} contain further examples of smooth resolutions of
cones over the $L^{p,q,r_1,\ldots,r_{n-1}}$ Einstein-Sasaki metrics.
This forms the subject of the present paper.  Our conclusions are that
in six dimensions, there are no further examples going beyond the
resolution of the cone over $Y^{2,1}$ that was obtained in
\cite{ootyas}.  In particular, it does not seem to be possible to
obtain resolutions of cones over any of the other $L^{pqr}$
Einstein-Sasaki spaces, within the framework of the local metrics
found in \cite{chlupo3}.  Thus although the local six-dimensional
metrics in \cite{chlupo3} in general have cohomogeneity 3, it seems
that one only obtains complete and non-singular examples under the
specialisation to cohomogeneity 2.

   However, if we go beyond six dimensions we find large classes of
new examples of complete and non-singular metrics on resolved cones
over inhomogeneous Einstein-Sasaki spaces.  Motivated by our detailed
analysis in six dimensions, it is natural to focus attention on the
specialisation of the BPS metrics in \cite{chlupo3} to cohomogeneity
2.  (In general, in $D=2n$ dimensions, the BPS metrics in
\cite{chlupo3} have cohomogeneity $n$.)  The special cases of
cohomogeneity-2 Kerr-NUT-AdS metrics, and their BPS limits, were
obtained previously in \cite{chlupo1}, and it is in fact more
convenient to take these results as the starting point for the
investigation of the global regularity issues.

     The organisation of this paper is as follows.  In section 2, we
review the construction of local Ricci-flat K\"ahler metrics that were
obtained in \cite{chlupo1} by taking a BPS limit of the
cohomogeneity-2 Kerr-NUT-AdS that were also constructed in that paper,
giving details that will be useful for our subsequent global analysis.
In section 3, we then set up the machinery for studying the
circumstances under which these local metrics can extend smoothly onto
complete and non-singular manifolds.  We do this by using the method
that was developed in \cite{cvlupapo1,cvlupapo2}, which involves
checking the compatibility between the periodicities of the
translations generated by the various Killing vectors whose lengths go
to zero on degenerating hypersurfaces.  The construction in
\cite{chlupo1} yields, in the BPS limit, Ricci-flat K\"ahler metrics
of dimension $D=2n+4$ built over a $2n$-dimensional Einstein-K\"ahler
base metric $d\Sigma_{n}^2$.  We find that we can obtain complete and
non-singular spaces if the Einstein-K\"ahler base is chosen to be a
product of $n$ $S^2$ factors, or to be $\CP^n$, a product of $m$ $S^2$
factors and $\CP^{n-m}$, or various other products involving complex
projective spaces as factors.  The special case with a single $S^2$
manifold forming the base corresponds to the example discussed in
\cite{ootyas}.

    In section 4, we discuss the topology of the manifolds on which
the smooth metrics are defined.  They are of the form of an $\R^2$
bundle over a manifold that is itself an $S^2$ bundle over a the
$2n$-dimensional Einstein-K\"ahler base manifold.

    In section 5, we present a discussion of the more general
cohomogeneity-$n$ BPS limits of $2n$-dimensional Kerr-NUT-AdS metrics
that were obtained in \cite{chlupo3}.  Although we have not found any
further smooth resolutions amongst these more general metrics, it is
nonetheless of interest to study them in some detail, since their
local interpretation as deformed cones over the
$L^{p,q,r_1,\cdots,r_{n-1}}$ Einstein-Sasaki metrics was not
considered in \cite{chlupo3}.  In the process of doing this, we
incidentally obtain new, and rather elegant, expressions for the
local $L^{p,q,r_1,\cdots,r_{n-1}}$ Einstein-Sasaki metrics that were
first constructed in a rather different formalism in \cite{cvlupapo1}.
We also present a detailed discussion of the six-dimensional case,
showing how a global analysis leads to the conclusion that there are
no further smooth resolutions of cones over 5-dimensional
Einstein-Sasaki spaces within this class that go beyond the $Y^{2,1}$
example in \cite{ootyas}.

   The paper ends with conclusions in section 6.

\section{The Local Ricci-Flat K\"ahler Metrics}

   Our starting point is the class of local Ricci-flat K\"ahler
metrics that were obtained in \cite{chlupo1} by taking a BPS limit of
the cohomogeneity-2 Kerr-NUT-AdS metrics that were constructed in that
paper.  (They can also obtained from some Einstein-K\"ahler metrics
given in \cite{lupova}, by taking a limit where they become Ricci
flat.)  They have dimension $D=2n+4$, and are given by
\bea
d\td s^2 &=& \fft{x-y}{4 X} dx^2 + \fft{x-y}{4Y} dy^2 +
\fft{(x-\alpha)(\alpha -y)}{\alpha} d\Sigma_n^2\nn\\
&&+\fft{X}{x-y} \Big[d\tau + \fft{\alpha-y}{\alpha}(d\psi + A)\Big]^2
+\fft{Y}{x-y}\Big[d\tau - \fft{x-\alpha}{\alpha} (d\psi +
A)\Big]^2\,,\nn\\
X&=&x(x-\alpha) + \fft{2\mu}{(x-\alpha)^n}\,,\qquad
Y=y(\alpha-y) - \fft{2\nu}{(\alpha-y)^n}\,,\label{coho2genmetric}
\eea
where $d\Sigma_n^2$ is a metric on a $2n$-dimensional
Einstein-K\"ahler space $Z$, satisfying $R_{ab}=2(n+1)\, g_{ab}$, with
K\"ahler form $J= \ft12 dA$.

   There is a complete symmetry between the coordinates $x$ and $y$ in
the local metric.  However, when we consider the circumstances under
which we can obtain globally non-singular spaces, we shall find that
one of the two coordinates should be interpreted as a non-compact
radial variable, ranging from infinity to some specific finite value,
while the other coordinate should be interpreted as a compact
variable, analogous to a latitude coordinate on a sphere, ranging
between two finite values.  For definiteness, and without loss of
generality, we shall choose to take $x$ as the radial variable, and
$y$ as the coordinate ranging over the finite interval.  Specifically,
we shall take
\be
\alpha<x_0 \le x \le \infty\,,\qquad
\hbox{or}\qquad -\infty \le x\le x_0\,,
\ee
where $x_0$ is the largest (or, respectively, smallest) real root of
$X(x)$, and
\be
y_1 \le y \le y_2 \,,
\ee
where $y_1$ and $y_2$ are two adjacent real roots of $Y(y)$,
satisfying the conditions
\be
 y_1 < y_2 < \alpha\,.
\ee
The constant $\alpha$ can be fixed to any specific non-zero value 
without any loss of generality, and so from now on we shall take
\be
\alpha=1\,.
\ee
When these conditions are satisfied, the power-law curvature
singularities at $x=y$, $x=\alpha$ and $y=\alpha$ can all be avoided,
provided that $x_0>\alpha=1$ in the case that $x_0$ is the largest
real root of $X(x)$, or that $x_0<y_1$ in the case that $x_0$ is the
smallest real root of $X(x)$.  Note that in this latter case, for
which $x$ lies in the range $-\infty\le x \le x_0$, the metric $d\td
s^2$ will be negative definite, and so we should take $ds^2=-d\td s^2$
as the positive-definite Ricci-flat metric.
    
   We shall consider cases where the Einstein-K\"ahler base manifold
is taken to be the complex projective space $\CP^n$, or products of
$S^2$ factors, or products of $S^2$ factors and complex projective
spaces.  For the subsequent global analysis, it will be useful to
present a convenient coordinatisation of the standard Fubini-Study
metric on $\CP^n$.  In terms of the standard inhomogeneous complex
coordinates $\zeta^i$, the Fubini-Study metric and the potential $A$
are given by
\bea
d\Sigma_n^2 &=& f^{-1}\, \sum_{i=1}^n d\bar\zeta^i\, d\zeta^i -
    f^{-2}\, |\omega|^2\,,\nn\\
A &=& \ft{\im}{2}\, f^{-1}\, (\omega-\bar\omega)\,,\nn\\
f &=& 1 +\sum_{i=1}^n |\zeta^i|^2\,,\qquad 
  \omega = \sum_{i=1}^n \zeta^i\, d\bar\zeta^i\,.
\eea
We shall find it convenient to parameterise the $\zeta^i$ in terms of
real coordinates $r$, $\mu_i$ and $\phi_i$ according to
\bea
\zeta^i &=& \mu_i\, \tan\chi\, 
            e^{\im\, \phi_i}\,, \qquad \sum_{i=1}^n \mu_i^2=1\,,\nn\\
 0\le\chi\le \ft12\pi\,, &&\qquad 0\le\phi_i <2\pi\,,\qquad 
   0\le \mu_i\le 1\,.
\eea
in terms of which the $\CP^n$ metric and potential $A$ become
\bea
d\Sigma_n^2 &=& d\chi^2 + \sin^2\chi\, \Big[ \sum_{i=1}^n (d\mu_i^2 + 
     \mu_i^2\, d\phi_i^2) - 
  \sin^2\chi\, \Big(\sum_{i=1}^n \mu_i^2\, d\phi_i\Big)^2\Big]\,,\nn\\
A &=& \sin^2\chi\, \sum_{i=1}^n \mu_i^2\, d\phi_i\,.\label{cpnmet}
\eea
The metric is Einstein-K\"ahler, with $R_{ab} = 2(n+1)\, g_{ab}$. 

    When considering the case where $d\Sigma_n^2$ is instead the
Einstein-K\"ahler metric on the product of $n$ copies of $S^2$, the
metric and the potential $A$ can be taken to be
\bea
d\Sigma_n^2 &=& \fft1{2(n+1)}\, 
  \sum_{i=1}^n (d\theta_i^2 + \sin^2\theta_i\, d\phi_i^2)
\,,\nn\\
A&=& \fft1{n+1}\, \sum_{i=1}^n \cos\theta_i\, d\phi_i\,.\label{s2nmet}
\eea
Note that we have scaled the metric so that it has the 
canonical normalisation $R_{ab}= 2(n+1)\, g_{ab}$, as required in the
construction of the Ricci-flat metrics (\ref{coho2genmetric}).

\section{Complete Non-Singular Calabi-Yau Spaces}\label{cohom2sec}

   In this section we shall show how, by choosing the parameters in
the local Ricci-flat K\"ahler metrics (\ref{coho2genmetric})
appropriately, we can obtain metrics that extend smoothly onto
complete and non-singular manifolds.  We shall discuss various choices
for the Einstein-K\"ahler base $d\Sigma_n^2$ in the following
subsections.

\subsection{$\CP^n$ base space}

    In this case we consider the local Ricci-flat K\"ahler metrics
(\ref{coho2genmetric}) with $d\Sigma_n^2$ and $A$ chosen to be those
of the Fubini-Study metric on $\CP^n$, as given by (\ref{cpnmet}).

   For generic values of the coordinates, the metric
(\ref{coho2genmetric}) is non-singular.  It becomes singular on
``degeneration surfaces,'' where one or more of the metric
coefficients goes to zero.  These singularities can occur in the
$\CP^n$ base metric (\ref{cpnmet}), where one or more of the latitude
coordinates $\mu_i$ goes to zero, or where $\chi$ goes to 0 or
$\ft12\pi$.  Singularities can also occur in the total metric
(\ref{coho2genmetric}) where the functions $X(x)$ or $Y(y)$ go to
zero.  Thus these singularities occur at the inner end $x=x_0$ of the
range of the radial variable $x$, and at the two endpoints $y=y_1$ and
$y=y_2$ of the ``latitude'' coordinate $y$.  At all these degeneration
surfaces, some angular variable parameterising translations around the
surface becomes degenerate.

   In order for the local metric to extend smoothly onto a complete
and non-singular manifold, it is necessary that at all these
degeneration surfaces, the periodicities of the relevant collapsing
angular variables be such that the metric locally has the form of a
regular origin of polar or hyperspherical polar coordinates.  The
requirement that the periodicities needed to avoid conical
singularities at the various degeneration surfaces should all be
compatible is highly non-trivial, and it places severe restrictions on
the possible choices of periods and metric parameters.  A convenient
way to study these compatibility conditions was developed in
\cite{cvlupapo1}.  It consists of identifying the specific Killing
vectors whose lengths go to zero at the various degeneration surfaces,
in each case normalising them so that they generate precisely $2\pi$
translations around their respective surfaces.  There will exist
linear relations among the set of Killing vectors so obtained, and
compatibility will require that the coefficients in these linear
relations must be sets of coprime integers.  The task is then to check
whether there exist choices of coprime coefficients that are mutually
compatible at all the degeneration surfaces, and that are consistent
with the disposition of zeros in the metric functions $X(x)$ and
$Y(y)$.  Lower-dimensional degeneration surfaces, for example where
$X(x)$ and $Y(y)$ vanish simultaneously, must also be considered.

  In all cases we normalise each Killing vector $\ell$ so that it
generates a $2\pi$ translation at its respective degeneration surface
where $\ell^2\equiv \ell^\a \, \ell^b\, g_{ab}=0$.  This is achieved
by scaling it so that its ``Euclidean surface gravity'' $\kappa$,
defined by taking the limit as one approaches the degeneration surface
of
\be
  \kappa^2 = \fft{g^{ab}\, \del_a(\ell^2)\, \del_b(\ell^2)}{4 \ell^2}\,,
\ee
is equal to unity.\footnote{The {$\mathfrak{\hbox{ur}}$} example of
this kind is the Killing vector $\del/\del\phi$ at the origin of polar
coordinates in the metric $ds^2= dr^2 + r^2 d\phi^2$.}  For the case
of the metric (\ref{coho2genmetric}) with $\CP^n$ base, we find the
following normalised Killing vectors that vanish on degeneration
surfaces:
\bea
x=x_0:&& \ell_0= \fft{1}{[1- (n+2) x_0]}\, 
               \fft{\del}{\del\psi} +
    \fft{(x_0-1)}{[1-(n+2) x_0]}\, \fft{\del}{\del\tau}\,,\nn\\
y=y_i: && \ell_i= \fft{1}{[1- 
      (n+2) y_i]}\, \fft{\del}{\del\psi} +
    \fft{(y_i-1)}{[1-(n+2) y_i]}\, \fft{\del}{\del\tau}\,,\quad
i=1,2\,,\nn\\
\chi=\ft12\pi: &&
  \ell_3= \fft{\del}{\del\psi} - \sum_{i=1}^n \fft{\del}{\del\phi_i}
\,,\nn\\
\mu_i=0:&& \ell_{i+3} = \fft{\del}{\del\phi_i}\,,\qquad 1\le i\le n\,.
\label{kvs}
\eea

   As $|x|$ tends to infinity, the metric (\ref{coho2genmetric})
asymptotically approaches a cone over a $(2n+3)$-dimensional
Einstein-Sasaki space, contained within the class constructed in
\cite{cvlupapo1,cvlupapo2}.  (We shall show this in general in section
4.)  Thus before we address the new question of the resolution of the
singularity of the cone metric at the innermost value of $x$, we can
first consider the regularity conditions on the level surfaces at
constant $x$.  As $|x|$ tends to infinity this will just amount to
revisiting the regularity conditions for Einstein-Sasaki spaces
considered in \cite{cvlupapo1} (except that for now we are restricting
our attention to the special cases where the Einstein-Sasaki spaces
have cohomogeneity 1, which were also studied in \cite{gmsw2,clpv}.)
An important new feature is to note that the conditions for regularity
on the $(2n+3)$-dimensional level surfaces at fixed $x$ are {\it
independent of the value of $x$}, and thus regularity at large $x$
ensures regularity on all the non-degenerate level surfaces as $x$
approaches its innermost value $x=x_0$.  Finally, we can then address
the question of regularity on the degeneration surface at $x=x_0$
itself.  The requirements that this be regular and also compatible
with the conditions needed for regularity on the surfaces away from
$x=x_0$ are highly restrictive, admitting only very few consistent
solutions.  For example, in six dimensions it is these last
requirements which show that only for the case of $Y^{2,1}$ can one
obtain a smooth resolution of the cones over $Y^{p,q}$ Einstein-Sasaki
spaces \cite{ootyas}, within the class of metrics considered here.

    Starting, then, by considering the level surfaces away from $x=x_0$,
we note from (\ref{kvs}) that there must exist a linear relation
\be
p\ell_1 +q \ell_2 + \sum_{i=3}^{n+3} r_i\, \ell_i=0\label{linrel1}
\ee
among the Killing vectors.  As discussed in \cite{cvlupapo1} the
coefficients must be rationally related (since otherwise a linear
combination could generate a closed periodic translation through an
arbitrarily small length, implying a singularity in the manifold), and
thus by scaling we can take the coefficients to be coprime integers,
$p$, $q$, $r_i$.  From the detailed expressions for the Killing
vectors in (\ref{kvs}), we see from (\ref{linrel1}) that we must have
\be
r_3=r_4=\cdots=r_{n+3} = \fft{p+q}{n+1}\,,\label{eq1}
\ee
and also
\be
\fft{p}{q} + \fft{(1-y_2)[1-(n+2) y_1]}{(1-y_1)[1-(n+2)y_2]}=0\,.
\label{eq2}
\ee
Since $y_1$ and $y_2$ are roots of $Y(y)=0$, where $Y$ is given in
(\ref{coho2genmetric}), we have
\be
y_1\, (1-y_1)^{n+1} - y_2\, (1-y_2)^{n+1} = 0\,.\label{eq3}
\ee
Equations (\ref{eq2}) and (\ref{eq3}) determine $(y_1,y_2)$.

     The above considerations determine the structure of the level
surfaces away from the degeneration at the innermost radius $x=x_0$.
As we remarked previously, an important point in the analysis above is
that the compatibility conditions are independent of $x$, and thus the
conditions for regularity on the level surfaces are the same at
arbitrary values of $x\ne x_0$ as they are in the asymptotic region
where $|x|$ goes to infinity.  Thus the above analysis is essentially
equivalent to one given in \cite{cvlupapo1}.

   We now turn to the consideration of the degeneration at $x=x_0$.
 From (\ref{kvs}) it can be seen that there is a linear relation among
 the three Killing vectors $\ell_0$, $\ell_1$ and $\ell_2$, which we
 can express as
\be
n_0\, \ell_0 + n_1\, \ell_1 + n_2\, \ell_2 =0\label{nl}
\ee
for coprime integers $n_0$, $n_1$ and $n_2$.  Since there exists a
lower-dimensional hypersurface on which $\ell_1$ and $\ell_0$ both
have zero length simultaneously (and likewise a hypersurface where
$\ell_2$ and $\ell_0$ have zero length simultaneously), regularity
also imposes the stronger requirements that $n_1=n_2=1$. (See
\cite{cvlupapo1,cvlupapo2} for a more detailed exposition of
regularity considerations of this type.)  From the detailed
expressions for the Killing vectors in (\ref{kvs}), we see that
(\ref{nl}) then implies $n_0=-2$ and
\be
x_0= \fft{2(n+2) \, y_1\, y_2 - y_1 - y_2}{(n+2)(y_1+y_2) -2}\,.\label{eq4}
\ee
Note that in order for $y_1,y_2$ and $x_0$ to lie in the appropriate
range discussed in section 2, it is necessary that $p$ and $q$ be
positive integers.

      To summarise, the general linear relation
\be
\sum_{i=0}^{n+3}n_i\, \ell_i=0
\ee
for the zero-length Killing vectors imply four 
inequivalent conditions, namely
\bea
n_0=0:&& n_1=p\,,\quad n_2=q\,,\qquad n_i=\fft{p+q}{n+1}\,,\nn\\
n_1=1:&& n_0=2(n+1)p\,,\quad n_2=-(n+1)(p-q)\,,\quad
n_i=p+q\,,\quad i=3,4,\ldots n+3\,,\nn\\
n_2=1:&& n_0=2(n+1)q\,,\quad n_1=-(n+1)(p-q)\,,\quad
n_i=p+q\,,\quad i=3,4,\ldots n+3\,,\nn\\
n_i=0:&& n_0=-2\,,\quad n_1=1\,,\quad n_2=1\,,\quad
i=3,4,\ldots n+3\,.
\eea
The second and third cases imply the conditions
\be
\gcd[2(n+1)p, (n+1)(p-q), p+q]=
\gcd[2(n+1)q, (n+1)(p-q), p+q]=p+q\,,\label{gcdcon}
\,.
\ee
where gcd denotes the greatest common divisor.

   The condition (\ref{gcdcon}) imply that the pair of integers
$(p,q)$ can lie in either of two permitted classes.  One can either
have all the positive odd integer pairs such that $p+q=2(n+1)$, or
else one can have all the positive integer pairs such that $p+q=n+1$.
Thus for $\CP^0$ (\ie a zero-dimensional base space in
(\ref{coho2genmetric})), we can just have $(p,q)=(1,1)$.  This
corresponds to the Eguchi-Hanson metric.  For $\CP^1$, we can have
either $(p,q)=(1,1)$, corresponding to the cohomogeneity-1 resolution
of the cone over the homogeneous space $T^{1,1}/\Z_2$, or else
$(p,q)=(1,3)$.  This latter choice gives rise to the resolved cone
over $Y^{2,1}$, which was obtained in \cite{ootyas}.  For $\CP^2$, we
can have $(p,q)=(1,5)$, $(1,2)$, or $(3,3)$; and for $\CP^3$, we can
have $(p,q)=(1,7)$, $(1,3)$, $(3,5)$, or $(2,2)$.  In all cases, the
solution reduces to cohomogeneity 1 if $p=q$.

   For each allowed integer pair $(p,q)$, one then solves equations
(\ref{eq2}) and (\ref{eq3}) for $y_1$ and $y_2$, and then equation
(\ref{eq4}) for $x_0$.\footnote{Of course, as can be seen from the
expressions for $X(x)$ and $Y(y)$ in (\ref{coho2genmetric}), once
these roots are determined, the constants $\mu$ and $\nu$ are
determined.}  It is then necessary to check that the two roots $y_1$
and $y_2$ of $Y(y)$ are indeed adjacent, satisfying $y_1<y_2<1$, and
that the root $x_0$ of $X(x)$ is indeed the largest root, with
$x_0>y_2$ (and with $x$ ranging from $x_0\le x\le\infty$) or else
$x_0$ is the smallest root, with $x_0<y_1$ (and with $-\infty \le x\le
x_0$).  In fact in all the cases we have examined, one finds $x_0$ is
negative, and it is the smallest root.  With $-\infty\le x\le x_0$ the
metric $d\td s^2$ in (\ref{coho2genmetric}) is then negative definite,
and so $ds^2=-d\td s^2$ is a positive-definite Ricci-flat K\"ahler
metric on a complete and non-singular manifold.

  Some explicit examples are as follows.  In each case, we shall
consider only the non-trivial cohomogeneity-2 examples that arise when
$p\ne q$.  For the base $\CP^1$, we have only $(p,q)=(1,3)$, and this
implies
\be
y_1= \ft1{12}(5-\sqrt{13})\,,\qquad y_2= \ft1{12}(11-\sqrt{13})\,,\qquad
 x_0= -\ft13(1+\sqrt{13})\,.
\ee
This case coincides precisely with the six-dimensional resolution of
the cone over $Y^{2,1}$ that was found in \cite{ootyas}.  Note that
here, and in all the other examples, the radial variable $x$ runs from
$x_0$ to $-\infty$, and the positive-definite metric $ds^2$ is given
by (\ref{coho2genmetric}) with $ds^2 =-d\td s^2$.

  For $\CP^2$, we can have $(p,q)=(1,5)$ or $(1,2)$, which imply
\bea
(p,q)=(1,5):&&
  y_1 = 0.0311124\,,\qquad  y_2 = 0.647818\,,\qquad x_0=-0.72331\,,\nn\\
(p,q)=(1,2):&& 
y_1 = 0.119655\,,\qquad  y_2 = 0.421351\,,\qquad x_0= -0.839339\,.
\eea
(In these higher-dimensional examples the solutions for $y_1$ and $y_2$
arise as the roots of cubic or higher-order equations, and we shall
just give numerical values in these cases.)   

  For $\CP^3$ we can have:
\bea
(p,q)=(1,7):&& y_1 = 0.00922175\,, \qquad y_2 =0.659267\,,\qquad
  x_0= -0.452677\,,\nn\\
(p,q)=(1,3):&& y_1 = 0.0477244\,, \qquad y_2 = 0.459358\,,\qquad
  x_0= -0.537637\,,\nn\\
(p,q)=(3,5):&& y_1 = 0.112054\,,\qquad  y_2 = 0.313348\,,\qquad
  x_0=-0.584858\,.
\eea

\subsection{$(S^2)^n$ base space}

In this and subsequent subsections, we shall consider the cases where
$d\Sigma_n^2$ is a direct sum of various $\CP$ metrics.  One simple
case is to take $n$ 2-spheres $S^2$.  The metric is given by
(\ref{s2nmet}).  The zero-length Killing vectors are given by
\bea
x=x_0:&& \ell_0 = \fft{n+1}{[1 - (n+2) x_0]} \,\fft{\del}{\del\psi} +
\fft{x_0-1}{[1 - (n+2) x_0]} \,\fft{\del}{\del\tau}\,,\nn\\
y=y_1:&& \fft{n+1}{[1 - (n+2) y_1]} \,\fft{\del}{\del\psi} +
\fft{y_1 -1}{[1 - (n+2) y_1]} \,\fft{\del}{\del\tau}\,,\nn\\
y=y_2:&& \fft{n+1}{[1 - (n+2) y_2]} \,\fft{\del}{\del\psi} +
\fft{y_2 -1}{[1 - (n+2) y_2]} \,\fft{\del}{\del\tau}\,,\nn\\
\theta_i=0:&& \ell_{2i+1} = \kdel{\psi} - \kdel{\phi_i}\,,\qquad
i=1,2,\ldots,n\,,\nn\\
\theta_i=\pi:&& \ell_{2i+2} = \kdel{\psi} + \kdel{\phi_i}\,,\qquad
i=1,2,\ldots,n\,. \label{ns2kvs}
\eea
In the asymptotic region where $x$ becomes large, the metric describes
a cone over a $(2n+3)$-dimensional Einstein-Sasaki space.  The
periodicity conditions on the angular coordinates of the
Einstein-Sasaki space can be determined by analysing the zero-length
Killing vectors $(\ell_1, \ell_2, \ldots\, \ell_{2n+2})$, as in the
previous subsection.  Since the $S^2$ factors enter symmetrically, and
$\ell_0, \ell_1, \ell_2$ do not involve the vectors $\del/\del\phi_i$,
it suffices to check the conditions associated with one of the $S^2$
factors.  Thus we need only check the consistency conditions
associated with the Killing vectors $(\ell_0,\ell_1,\ell_2,
\ell_3,\ell_4)$.  As discussed in the previous case, the linear
dependence of these vectors implies that there exists a relation
\be
n_0\, \ell_0 + n_1\, \ell_1 + n_2\, \ell_2 + n_3\, \ell_3 +
n_4\, \ell_4=0\,,\label{d2nellcon}
\ee
for a set of coprime integers $n_j$.  

    The conditions associated with the regularity of the level
surfaces at constant $x\ne x_0$ are determined by considering the
linear relation involving only $(\ell_1,\ell_2,\ell_3,\ell_4)$, and so
for this part of the analysis we set $n_0=0$.  We shall take the
remaining integers in (\ref{d2nellcon}) to be given by
$(n_1,n_2,n_3,n_4)=(\td p,\td q, \td r\, \td s)$.  From the
expressions for the $\ell_i$ we can conclude that
\be
\td s=\td r=\ft12 (\td p+\td q)\,,
\ee
and in addition, $y_1$ and $y_2$ need to satisfy a condition that is
in fact identical to the one that arose for the $\CP^n$ base, namely
(\ref{eq2}).  Furthermore, since $y_1$ and $y_2$ are roots of
$Y(y)=0$, they must also satisfy the condition (\ref{eq3}), as in the
case of the $\CP^n$ base.

    Having determined $y_1, y_2$, and hence $\nu$, in terms of
$(p,q)$, the nature of the spaces that form the constant-$x$ level
surfaces are fully determined.  It remains to investigate the
consistency conditions associated with the degeneration surface at
$x=x_0$.  It is evident from (\ref{ns2kvs}) that $(\ell_0, \ell_1,
\ell_2)$ are linearly dependent.  Taking $n_3=0=n_4$ in
(\ref{d2nellcon}), we see that $n_1\,\ell_1 +n_2\ell_2 + n_0\ell_0=0$
implies $n_0=-(n_1 + n_2)$.  Now, $\ell_0$ and $\ell_1$ can have zero
length simultaneously (at $x=x_0$, $y=y_1$) and similarly so can
$\ell_0$ and $\ell_2$ (at $x=x_0$, $y=y_2$). This implies that we must
have $n_1=1=n_2$, and hence $n_0=-2$.\footnote{If instead we chose
$n_1=-n_2=1$, and hence $n_0=0$, it would force $y_1=y_2$, which would
reduce the metric to cohomogeneity 1, which has been considered
previously.}  We can also solve for $x_0$; it is given by
$(\ref{eq4})$, which is again the same as in the $\CP^n$ case
considered earlier.

       Now we have only two non-trivial conditions left, namely
\bea
n_1=0:&& n_0=4p\,,\qquad n_2=-2(p-q)\,,\qquad n_3=n_4=p+q\,,\nn\\
n_2=0:&& n_0=4q\,,\qquad n_1=2(p-q)\,,\qquad n_3=n_4=p+q\,.
\eea
This implies that
\bea
\gcd[4p, p+q, 2(p-q)]=p+q\,,\qquad
\gcd[4q, p+q, 2(p-q)]=p+q\,,
\eea
The only positive integer solutions with $p\le q$ are 
$(p,q)=(1,1)$ and $(p,q)=(1,3)$.  The $p=q$ case reduces to
previously-considered cohomogeneity-1 examples, which are not of interest
to us in this paper.

       Thus we see that for each allowed integer pair $(p,q)$, the
determination of $y_1$, $y_2$ and $x_0$ depends only on the dimensions
of the base space, given by (\ref{eq2}), (\ref{eq3}) and
(\ref{eq4}). The allowed pairs $(p,q)$ are determined simply by
examining one of the $S^2$ factors; no further restrictions follow by
examining the other $S^2$ factors.

Some explicit examples of the cohomogeneity-2 metrics with
$(p,q)=(1,3)$ are
\bea
D=6:&& y_1=\ft1{12} (5-\sqrt{13})\,,\qquad
y_2=\ft1{12} (11-\sqrt{13})\,,\qquad
x_0=-\ft13 (1 + \sqrt{13})\,.\nn\\
D=8:&& y_1=0.0696131\,,\qquad y_2=0.525812\,,\qquad
x_0=-0.792761\,,\nn\\
D=10:&& y_1= 0.0477244\,,\qquad
y_2= 0.459358\,,\qquad
x_0= -0.537637\,.
\eea
The example in $D=6$ is the resolution of the cone over $Y^{2,1}$ that
was found in \cite{ootyas}.  In all cases $y_1$ and $y_2$ are indeed
both less than 1, which means there will be no singularity within the
interval $y_1\le y\le y_2$.  Since $x_0$ is negative we again, as in
the cases with the $\CP^n$ base, take $x$ to range from $-\infty$ to
$x_0$, which ensures that $X(x)$ remains non-negative, and approaches
zero only at the degeneration surface at $x=x_0$.  Furthermore, the
range of the $x$ coordinate does not overlap with $x=1$ or the range
of the $y$ coordinate.  Thus we obtain regular positive-definite
metrics $ds^2 =-d\td s^2$ that extend smoothly onto non-singular
manifolds.

\subsection{$(\CP^m)^s$ base space}

      The above analysis of $n$ copies of $S^2$ can be easily
generalised to $s$ copies of $\CP^m$, where $n=s\, m$. The formulae
to determine $y_1,y_2$ and $x_0$ are the same, given by (\ref{eq2}),
(\ref{eq3}) and (\ref{eq4}); these depend only on the total dimension
$2n$ of the base space.  The allowed $(p,q)$ values are the same as
those for the single copy of $\CP^m$, namely positive odd integer
pairs $(p,q)$ satisfying $p+q=2(m+1)$, and positiver integer pairs
satisfying $p+q=m+1$.

\subsection{$\CP^{m_1}\times \CP^{m_2}\times \cdots\times 
        \CP^{m_i}$ base space }

       If the metric $d\Sigma_n^2$ on the base space is taken to be a
direct sum of different $\CP^{m_j}$ metrics, the basic formulae
governing the allowed values of $y_1$, $y_2$ and $x_0$ are again given
by (\ref{eq2}), (\ref{eq3}) and (\ref{eq4}).  However, there is a set
of such restrictions on the allowed $(p,q)$ values associated with
each of the $\CP^{m_j}$ factors in the base space.  The final allowed
set of $(p,q)$ pairs is therefore given by the intersection of all the
allowed sets associated with each factor.  For example, let us
consider $S^2\times \CP^2$.  The allowed $(p,q)$ for $S^2$ are given
by $(1,3), (1,1)$, and those for $\CP^2$ are $(1,5), (1,2), (3,3)$.
There is no intersection, and hence the orbifold singularities cannot
be resolved in this case.  On the other hand, if we consider
$S^2\times \CP^3$, the $(p,q)$ pairs allowed by $\CP^3$ are $(1,7),
(1,3), (3,5),(2,2)$, and so we have an intersection with the $(1,3)$
allowed by $S^2$.  It follows that we can obtain a complete metric
with base space $S^2\times \CP^3$, provided that we take
$(p,q)=(1,3)$.

     In general, we may consider the case where the $2n$-dimensional
base space is the direct product $\CP^{m_1}\times \CP^{m_2}\times
\cdots\times \CP^{m_i}$, with $n=(m_1 + m_2 + \cdots m_i)$.  Let
$P_j=\{(p,q)\}$, the set of positive odd integer pairs satisfying
$p+q=2(m_j+1)$ together with the set of positive integer pairs
satisfying $p+q=m_j+1$.  We determine the set $P_j$ for each factor
$\CP^{m_j}$, for $1\le j\le i$.  The final allowed set of pairs
$(p,q)$ that gives complete metrics is given by the intersection of
all $P_j$.

   This leads to the following conclusion for our metrics: There can
be no more than two different types of $\CP^{m_j}$ spaces in the
product base space.  Thus, the most general allowed $2n$-dimensional
bases space of this type comprises a direct product of $s_1$ copies of
$\CP^{m_1}$ and $s_2$ copies of $\CP^{m_2}$, with $s_1\, m_1 + s_2\,
m_2=n$.  If $m_1=m_2$, it reduces to a situation that we discussed
above.  If $m_1$ and $m_2$ are not equal, then the allowed $m_1$ and
$m_2$ are given by $m_2=2m_1 + 1$.  The allowed $(p,q)$ the odd
integer pairs that satisfy $p+q=2(m_1+1)$.

\section{Topology}

    Having established the conditions under which we can obtain
complete and non-singular metrics, we may now consider the topology of
the manifolds on which they are defined.  This can be understood by
examining the geometry in the neighbourhood of the degeneration
surface at $x=x_0$.  This surface is usually referred to as a bolt.
Recalling that $x$ ranges from the negative value $x_0$ to $-\infty$,
we define $x-x_0= -\rho^2$, and so from (\ref{coho2genmetric}) the
metric $ds^2=-d\td s^2$ near $\rho=0$ approaches
\be
ds^2 \sim \fft{y-x_0}{1-(n+2) x_0} \Big(d\rho^2  +
\fft{(1-(n+2)x_0)^2}{(1-x_0)^2}
\rho^2\, (d\tau + \cA)^2 \Big) + ds_{\rm bolt}^2\,,
\ee
where
\bea
\cA &=& \Big(\fft{(1-x_0)(1-y)}{y-x_0} + \fft{(1-x_0)^2 Y}{
((n+2)x_0 -1) (y-x_0)^2}\Big) d\wtd \psi\,,\nn\\
ds_{\rm bolt}^2 &=& \fft{(1-x_0)^2 Y}{y-x_0} (d\wtd \psi + A)^2
+\fft{y-x_0}{4Y} dy^2 + (1-x_0)(1-y) d\Sigma_n^2\,,
\eea
with $\wtd \psi=\psi + \tau/(1-x_0)$.  It can be seen $\wtd \psi$ must
have an appropriate period in order to avoid conical singularities on
the bolt, at $y=y_1$ and $y=y_2$.  That a single choice of period
achieves regularity at both endpoints is automatically guaranteed by
virtue of our previous general analysis of the degeneration surfaces.
Specifically, it can be seen to follow from (\ref{eq2}) and
(\ref{eq4}); the period is given by
\be
\Delta\wtd \psi = \fft{2\pi (p+q)}{(n+1)(q-p)}\,.
\ee
The coordinates $y$ and $\wtd\psi$ therefore parameterise a 2-sphere,
and the bolt $B$ is described as an $S^2$ bundle over the
Einstein-K\"ahler base space $Z$ whose metric is $d\Sigma_n^2$.  The
total $(2n+4)$-dimensional space on which the metric $ds^2=-d\td s^2$
is defined is an $\R^2$ bundle over the $S^2$ bundle over the bolt
$B$.

         Note that in the case when $p=q$, corresponding to
cohomogeneity-1 metrics, the bolt metric becomes homogeneous, and is a
direct product of $S^2$ with the base space $Z$ whose metric is
$d\Sigma_n^2$.  In the case of $D=6$, for example, the metric is the
resolution of the cone over $T^{11}/\Z_2$ with an $S^2\times S^2$
bolt, rather than the deformation of the conifold over $T^{11}$
\cite{conifold} which has an $S^3$ bolt.

\section{Metrics of Higher Cohomogeneity}

    So far, we have restricted our attention to the cohomogeneity-2
Ricci-flat K\"ahler metrics that were obtained in \cite{chlupo1} as
BPS limits of the cohomogeneity-2 Kerr-NUT-AdS metrics that were also
constructed in that paper.  Subsequently, the construction of
Kerr-NUT-AdS metrics was generalised to higher cohomogeneity in
\cite{chlupo3}.  Specifically, in even dimensions $D=2n$, Kerr-NUT-AdS
metrics of cohomogeneity $n$ were constructed, with $(n-1)$
independent NUT parameters in addition to the mass and the $(n-1)$
independent rotation parameters \cite{chlupo3}.  A BPS limit was also
taken, yielding local $D=2n$ dimensional Ricci-flat K\"ahler metrics
of cohomogeneity $n$.  In this section, we shall study these metrics
in more detail, showing, in particular, how they can be viewed as
deformations of cones over the classes of Einstein-Sasaki spaces that
were constructed in \cite{cvlupapo1,cvlupapo2}.  In the simplest case
of $D=6$, we shall also give a global analysis of the cohomogeneity-3
metrics, showing that the only smooth resolutions of the cones over
$L^{pqr}$ Einstein-Sasaki spaces arise in the specialisation to
cohomogeneity 2, which leads back the single example with the
$Y^{2,1}$ space that was obtained in \cite{ootyas}.

\subsection{Cohomogeneity-$n$ Calabi-Yau metrics in $D=2n$}\label{gennsec}

    The Ricci-flat K\"ahler metrics of cohomogeneity $n$ in $D=2n$ 
dimensions that were obtained in \cite{chlupo3} are given by
\bea 
ds_{2n}^2 &=& \sum_{\mu=1}^n \Big[ \fft{U_\mu\, dx_\mu^2}{4X_\mu}
   +
\fft{X_\mu}{U_\mu} \Big(
\fft{\gamma}{x_{\mu}} d\tau - \sum_{i=1}^{n-1} \fft{W_i\,
d\phi_i}{\alpha_i-x_\mu}
\Big)^2\Big]\,,\nn\\
X_\mu &=& x_\mu \prod_{i=1}^{n-1} (\alpha_i - x_\mu) -
2\ell_\mu\,,\qquad U_\mu= {{\prod}'}^n_{\nu=1} (x_\nu-x_\mu)\,,\nn\\
W_i &=& \prod_{\nu=1}^n (\alpha_i -
x_\nu)\,,\qquad \gamma = \prod_{\nu=1}^n x_{\nu}\,,\label{d2nmet}
\eea
where the prime on a product symbol universally indicates that the
factor that vanishes is to be omitted.

    The $n$ coordinates $x_\mu$ enter in a completely symmetrical
fashion, and any one of them can be selected to play the role of a
radial variable, which takes values in an infinite range from infinity
to some finite value at the ``centre.''  For definiteness, we shall
choose $x_n$ as the non-compact radial variable, setting $x_n=r^2$.
The remaining $x_\alpha$ coordinates with $1\le \alpha \le n-1$ are
analogous to latitude coordinates on a sphere, and they range between
adjacent roots of the corresponding polynomials $X_\alpha$ that appear
in the metric.

    It is easily seen that at large $r$ the metric
(\ref{d2nmet}) takes the form
\be
ds_{2n}^2 \rightarrow dr^2 + r^2 ds_{2n-1}^2\,,
\ee
where $ds_{2n-1}^2$ is an Einstein-Sasaki metric, which is given by
\be
ds_{2n-1}^2 = \sum_{\beta=1}^{n-1}\Big[ \fft{\hat U_\beta}{4 X_\beta}\,
 dx_\beta^2 + \fft{X_\beta}{\hat U_\beta}\,
   \Big(\fft{\hat\gamma}{x_\beta}\, d\tau +
      \sum_{i=1}^{n-1}\fft{\hat W_i\, d\phi_i}{\alpha_i-x_\beta}\Big)^2
  \Big]  +
  \Big( \hat\gamma\, d\tau - \sum_{i=1}^{n-1} \hat W_i\, d\phi_i\Big)^2\,,
\label{es2nm1}
\ee
where
\be
\hat U_\alpha= {{\prod}'}_{\beta=1}^{n-1} (x_\beta-x_\alpha)\,,\qquad
\hat W_i= \prod_{\beta=1}^{n-1} (\alpha_i-x_\beta)\,,\qquad
\hat\gamma=\prod_{\beta=1}^{n-1} x_\beta\,,
\ee
and $X_\alpha$ is given by the expression for $X_\mu$ in (\ref{d2nmet})
with $\alpha=\mu$ for $1\le\alpha\le n-1$.
It is straightforward to show that the Killing vector
\be
K= \fft1{c}\, \fft{\del}{\del\tau} - \sum_{i=1}^{n-1} \fft1{\alpha_i\, B_i}\,
   \fft{\del}{\del\phi_i}
\ee
has unit length, $|K|^2=1$, where the constants $c$ and $B_i$ are given by
\be
c= \prod_{i=1}^{n-1} \alpha_i\,,\qquad
  B_i= {{\prod}'}_{j=1}^{n-1} (\alpha_i-\alpha_j)\,.
\ee

   In fact $K$ is the Reeb vector for the Einstein-Sasaki space.  It can
be written as $K=\del/\del\psi$ by defining new coordinates
$(\psi,\hat\phi_i)$ that are related to $(\tau, \phi_i)$ by
\be
\tau=\fft1{c}\, \psi\,,\qquad
      \phi_i= \hat\phi_i - \fft1{\alpha_i\, B_i}\, \psi\,.
\label{redefs}
\ee
The Einstein-Sasaki metric (\ref{es2nm1}) can then be written as
\be
ds_{2n-1}^2 = (d\psi+ {\cal A})^2 + d\bar s_{2n-2}^2\,,\label{base}
\ee
where $d\bar s_{2n-2}^2$ is an Einstein-K\"ahler metric, whose
K\"ahler form is given by $J= \ft12 d{\cal A}$.  This Einstein-K\"ahler
metric can be shown to be equivalent to the one introduced in
\cite{cvlupapo1}, which arose in the construction of Einstein-Sasaki
spaces $L^{p,q,r_1,\ldots,r_n}$ in arbitrary odd dimensions $(2n+1)$.

   It is straightforward to show that in terms of the new coordinates
defined in (\ref{redefs}), the Einstein-K\"ahler base metric $d\bar
s_{2n-2}^2$ can be written as
\be
d\bar s_{2n-2}^2 = \sum_{\beta=1}^{n-1} \Big[
\fft{\hat U_\beta}{4X_\beta}\,
  dx_\beta^2 + \fft{X_\beta}{\hat U_\beta}\, \Big( \sum_{i=1}^{n-1}
   \fft{\hat W_i\, d\hat \phi_i}{\alpha_i-x_\beta}\Big)^2 \Big]\,.
\ee
The Kaluza-Klein potential $\cA$ in (\ref{base}) is given by
\be
\cA= -\sum_{i=1}^{n-1} \hat W_i\, d\hat\phi_i\,.
\ee

   The full Ricci-flat K\"ahler metric (\ref{d2nmet}) 
itself simplifies considerably if written using the
coordinates $(\psi,\hat\phi_i)$ introduced in (\ref{redefs}), becoming
simply
\be
ds_{2n}^2 = \sum_{\mu=1}^n \Big[ \fft{U_\mu\, dx_\mu^2}{4X_\mu}
   +
\fft{X_\mu}{U_\mu} \Big(d\psi - \sum_{i=1}^{n-1} \fft{W_i\,
d\hat\phi_i}{\alpha_i-x_\mu} \Big)^2 \Big]\,.\label{d2nmet2}
\ee
It is useful also to record that the inverse of the metric
(\ref{d2nmet2}) is given by
\be
\Big(\fft{\del}{\del s}\Big)^2 = \sum_{\mu=1}^n \Big[
   \fft{4 X_\mu}{U_\mu}\, \Big(\fft{\del}{\del x_\mu}\Big)^2 +
     \fft{S_\mu}{X_\mu U_\mu}\, \Big( \fft{\del}{\del \psi} +
  \sum_{i=1}^{n-1} \fft{1}{B_i\, (\alpha_i-x_\mu)}\,
    \fft{\del}{\del\hat\phi_i}\Big)^2\Big]\,
\ee
where we define
\be
S_\mu= \prod_{i=1}^{n-1}(\alpha_i -x_\mu)^2\,.
\ee

   We can now give a general discussion of the Killing vectors that vanish
on degenerate surfaces.  The general Killing vector can be written as
\be
\ell = c_0\, \fft{\del}{\del\psi} +
   \sum_{i=1}^{n-1} c_i\, \fft{\del}{\del\hat\phi_i}\,.\label{KV}
\ee
Let $x_\mu^0$ be a root of the polynomial $X_\mu; \ie X_\mu(x_\mu)=0$ at
$x_\mu = x_\mu^0$.  We find that the Killing vector (\ref{KV}) vanishes
at $x_\mu=x_\mu^0$ if the constants $c_i$ are chosen to be
\be
c_i = - \fft{c_0}{(x_\mu^0 - \alpha_i)\, B_i}\,.
\ee
The surface gravity $\kappa$ on this degenerate orbit is then given by
\be
\kappa^2 = \fft{c_0^2\, [X'_\mu(x_\mu^0)]^2}{S_\mu(x_\mu^0)}\,,
\ee
where the prime denotes a derivative with respect to the argument.
The Killing vector is given its canonical normalisation, meaning that
it generates a closed $2\pi$ translation, by choosing $c_0$ so that
$\kappa^2=1$.

   For example, if we use the coordinate $x$ to denote $x_\mu$ with $\mu=1$,
then the Killing vector that vanishes at the root $x=x_0$ of the polynomial
$X_1(x)$ is given by (\ref{KV}) with
\be
c_i = -\fft{c_0}{(x_0-\alpha_i)\, B_i}\,.
\ee
The surface gravity is given by
\be
\kappa^2 = \fft{c_0^2 \, [X_1'(x)]^2}{\prod_{i=1}^{n-1} (x_0-\alpha_i)^2}\,.
\ee

   It is quite involved to give a general discussion of the conditions
arising from requiring regularity and compatibility at all the
degeneration surfaces for these metrics.  In the following subsection
we shall examine the case of six dimensions in some detail, and show
that in fact the conditions for regularity imply that the parameters
in the metric must be specialised so that the cohomogeneity reduces
from 3 to 2.  This leads back to the situation discussed in the
previous section, and thus we shall conclude that in six dimensions
the only smooth resolution of a cone over $L^{pqr}$ spaces that lies
within the class of metrics (\ref{d2nmet}) is the $Y^{2,1}$ example
found in \cite{ootyas}.

\subsection{Cohomogeneity-3 metrics in $D=6$}

     Here we consider the specialisation of the discussion in section
\ref{gennsec} to the case of six dimensions.  We shall denote the
three coordinates $x_\mu$ by $x$, $y$ and $z$ in this case, choosing
$z$ as the radial coordinate, and the two parameters $\alpha_i$ will
be denoted by $\alpha$ and $\beta$.  Thus we need to consider
degeneration surfaces at $x=x_1$, $x=x_2$, $y=y_1$, $y=y_2$ and
$z=z_0$.  The associated normalised Killing vectors that vanish on
these surfaces will be labelled $(\ell_0,\ell_1,\ell_2,\ell_3,
\ell_4)$ respectively.  From the discussion at the end of section
(\ref{gennsec}), it follows that these Killing vectors are given by
\crampest{
\bea
\ell_0&=&  \fft{1}{\alpha\beta - 2(\alpha + \beta) z_0 +
3z_0^2}\, \Big((\alpha-z_0)(\beta-z_0)\fft{\del}{\del\psi} +
  \fft{\beta-z_0}{\beta-\alpha}\fft{\del}{\del\hat\phi_1} +
  \fft{\alpha-z_0}{\alpha-\beta}\fft{\del}{\del\hat\phi_2} \Big)\,,\\
\ell_i&=&  \fft{1}{\alpha\beta - 2(\alpha + \beta) x_i +
3x_i^2}\, \Big((\alpha-x_i)(\beta-x_i)\fft{\del}{\del\psi} +
  \fft{\beta-x_i}{\beta-\alpha}\fft{\del}{\del\hat\phi_1} +
  \fft{\alpha-x_i}{\alpha-\beta}\fft{\del}{\del\hat\phi_2} \Big)\,,
\ i=1,2\nn\\
\ell_{i+2}&=&  \fft{1}{\alpha\beta - 2(\alpha + \beta) y_i +
3y_i^2}\, \Big((\alpha-y_i)(\beta-y_i)\fft{\del}{\del\psi} +
  \fft{\beta-y_i}{\beta-\alpha}\fft{\del}{\del\hat\phi_1} +
  \fft{\alpha-y_i}{\alpha-\beta}\fft{\del}{\del\hat\phi_2} \Big)\,,
\  i=1,2\nn
\eea}

   Note that the Killing vector whose length goes to zero on a given
degeneration surface has coefficients that depend only on $\alpha$,
$\beta$ and the limiting value of the coordinate that parameterises
the approach to that specific surface.  In particular, this means that
the analysis of the regularity of the level surfaces away from the
degeneration at $z=z_0$ is independent of $z$, and therefore the
regularity of the level surfaces at general $z\ne z_0$ is assured once
the regularity of the asymptotic Einstein-Sasaki space is established.
Thus, as in the cohomogeneity-2 metrics we discussed earlier in the
paper, the discussion of regularity can be broken into two parts;
first analysing the regularity of the non-degenerate level surfaces,
and then analysing the regularity at the degeneration surface $z=z_0$.

   To be more explicit, at the asymptotic region where $|z|$ goes to
infinity, the metric describes a cone over the Einstein-Sasaki metric
$L^{pqr}$.  The periodicity conditions for the azimuthal coordinates
$\psi, \hat\phi_1, \hat\phi_2$ are given by the analysis of the
linearly-related Killing vectors $\ell_1, \ell_2, \ell_3, \ell_4$,
namely\footnote{Note that the coprime integers $p,q,r,s$ we are using
here are not necessarily the same as those in \cite{cvlupapo1}.  When
we speak of the Einstein-Sasaki space $L^{pqr}$, we are using this as
a generic name, rather than implying a specific correlation of
parameters.}
\be
p\, \ell_1 + q\, \ell_2 + r\, \ell_3 + s\, \ell_4=0\,.\label{pqrscond}
\ee
As mentioned above, these four
Killing vectors are independent of $z$, and so the deformed 
$L^{pqr}$ level surfaces at constant $z\ne z_0$ will also be 
regular if (\ref{pqrscond}) holds.

   Now, it remains to investigate the degeneration surface at $z=z_0$.
To do this, we should therefore include the associated Killing vector
$\ell_0$ in the linear relation.  In general, we shall have
\be
n_0\, \ell_0 +n_1\, \ell_1 + n_2\, \ell_2 + n_3\, \ell_3 +
n_4\, \ell_4=0\,.\label{5nrel}
\ee
The vector space is three-dimensional, and so this equation implies
three relations.  A particularly simple one is obtained by taking the
coefficient of $\del/\del\psi$ plus $\alpha(\alpha-\beta)/3$ times the
coefficient of $\del/\del\hat\phi_1$ plus $\beta(\beta-\alpha)/3$
times the coefficient of $\del/\del\hat\phi_2$.  This implies
\be
n_0+n_1+n_2+n_3+n_4=0\,.
\ee
Since the vector space is only three-dimensional we can consider
specialisations of (\ref{5nrel}) where any one of $n_1$, $n_2$, $n_3$ or
$n_4$ is set to zero.  We shall consider two such specialisations.

   First, consider the case where $n_4=0$.  This implies that $n_0+n_1
+ n_2 + n_3=0$.  Furthermore, it should be noted that $\ell_0$,
$\ell_1$ and $\ell_3$ can all have zero length simultaneously (at
$z=z_0$, $x=x_1$ and $y=y_1$).  Likewise, $\ell_0$, $\ell_2$ and
$\ell_3$ can all have zero length simultaneously.  This implies (see
\cite{cvlupapo1,cvlupapo2} for further discussion of such
multiply-degenerate surfaces) that
\be
\gcd(n_0, n_1, n_3)=n_2\,,\qquad
\gcd(n_0,n_2,n_3)=n_1\,.
\ee
Since $(n_0,n_1,n_2,n_3)$ are coprime, this means that $n_1=n_2=1$.

   The same argument goes through if we instead choose $n_3=0$.
Denoting the remaining coprime integers by $n_0'$, $n_1'$, $n_2'$ and
$n_4'$ in this case, we must have $n_1'=n_2'=1$.  Now, the set of
three relations following from (\ref{5nrel}) for these two cases can
be solved respectively for $(n_1,n_2,n_3)$ in terms of $n_0$, and for
$(n_1',n_2',n_4')$ in terms of $n_0'$.  Calculating the ratio
$(n_1/n_2)/(n_1'/n_2')$ (which we have just shown must equal 1 for
regularity), and subtracting 1, we obtain the simple result
\be
(x_1-x_2)(y_1-y_2)=0\,.
\ee
Thus to obtain regular spaces, we must either have $x_1=x_2$ or $y_1=y_2$.
This leads back to the cohomogeneity-2 metrics that we have already 
studied earlier in the paper.\footnote{For example, if $x_1=x_2$, which means
the range of the $x$ coordinate is pinched down to zero, one can obtain 
a non-singular metric by setting $x_1= x_0-\epsilon$, 
$x_2=x_0+\epsilon$, $x= x_0 - \epsilon\cos\theta$ and sending $\epsilon$
to zero.  The coordinate $\theta$ then parameterises the latitude of a
(homogeneous) round 2-sphere, and the cohomogeneity of the metric is
reduced from 3 to 2.}

   The upshot of the above discussion is that in six dimensions, there
are no non-singular cohomogeneity-3 spaces contained within the class
(\ref{d2nmet}), and thus the cohomogeneity-2 example with $Y^{2,1}$
level surfaces found in \cite{ootyas} is the only non-singular example
beyond cohomogeneity 1.  It is of interest to study the situation in
higher dimensions.

\section{Conclusions}

    In this paper, we have generalised a recent result obtained in
\cite{ootyas}, where a smooth resolution of the Calabi-Yau cone over
the five-dimensional Einstein-Sasaki space $Y^{2,1}$ was constructed.
The construction in \cite{ootyas} involved specialising the parameters
in a general class of Ricci-flat K\"ahler metrics that were obtained
in \cite{chlupo1,chlupo3} as BPS limits of the Kerr-NUT-AdS metrics
that were also constructed in those papers.  The same Ricci-flat
K\"ahler metrics also arise as a limit, in which the cosmological
constant is sent to zero, of Einstein-K\"ahler metrics found in
\cite{lupova}.

   Our generalised analysis showed that we can obtain non-singular
resolutions of Calabi-Yau cones over Einstein-Sasaki spaces in all
even dimensions $D\ge 6$, by specialising the parameters in the
metrics obtained in \cite{chlupo1,chlupo3}.  In all cases, the
topologies of the associated smooth manifolds are of the form of an
$\R^2$ bundle over a base space that is itself an $S^2$ bundle over an
Einstein-K\"ahler manifold $Z$.  We examined in particular the cases
where the Einstein-K\"ahler manifold $Z$ is taken to be a product of
complex projective spaces.  Details of our results are contained in
section \ref{cohom2sec}.

   All the non-singular resolutions that we found have cohomogeneity
2, meaning that the level surfaces at constant radius are themselves
of cohomogeneity 1.  These level surfaces asymptotically approach
Einstein-Sasaki spaces at large radius.  We also examined the more
general situation of the higher-cohomogeneity Ricci-flat K\"ahler
metrics obtained in \cite{chlupo3}.  In dimension $D=2n$, these have
cohomogeneity $n$. We showed that the level surfaces are asymptotic to
the general classes of Einstein-Sasaki metrics that were constructed
in \cite{cvlupapo1,cvlupapo2}.  However, a detailed analysis of the
situation in dimension $D=6$ shows that although in general these
cohomogeneity-3 metrics describe Calabi-Yau cones over the $L^{pqr}$
Einstein-Sasaki metrics, the conditions for the absence of conical
singularities can only be satisfied under the specialisation to
cohomogeneity 2.  This leads to the conclusion that within this class
of six-dimensional metrics, only the Calabi-Yau cone over $Y^{2,1}$
admits a non-singular resolution.  It is of interest to study the
situation in higher dimensions, to see whether the non-singular
resolutions we have constructed in this paper are the only ones within
the class of metrics found in \cite{chlupo3}.

\section*{Acknowledgements}

    We thank Wei Chen, Mirjam Cveti\v c and Gary Gibbons for helpful
discussions.  C.N.P. is grateful to the Cosmology and Gravitation
group in the Centre for Mathematical Sciences, Cambridge, for
hospitality during the course of this work.


\begin{thebibliography}{99}

\bm{egha} T. Eguchi and A.J. Hanson,
{\it Asymptotically flat selfdual solutions to Euclidean gravity},
Phys. Lett. {\bf B74}, 249 (1978).

\bm{begipapo} V.A. Belinsky, G.W. Gibbons, D.N. Page and C.N. Pope,
{\it Asymptotically Euclidean Bianchi IX metrics in quantum gravity},
Phys. Lett. {\bf B76}, 433 (1978).

\bm{berber} L. Berard-Bergery, {\it Quelques examples de varietes
riemanniennes completes non compactes a courbure de Ricci positive},
C.R.\ Acad.\ Sci,\ Paris, Ser.\ 1302, 159 (1986).

\bm{pagpop}
D.N. Page and C.N.~Pope,
{\it Inhomogeneous Einstein metrics on complex line bundles},
Class.\ Quant.\ Grav.\  {\bf 4}, 213 (1987).

\bm{conifold} P. Candelas and X.C. de la Ossa,
{\it Comments on conifolds,}
Nucl.\ Phys.\ {\bf B342}, 246 (1990).

\bm{brysal} R.L. Bryant and S. Salamon, {\it On the construction
of some complete metrics with exceptional holonomy}, Duke Math. J.
{\bf 58}, 829 (1989).

\bm{gibpagpop} G.W. Gibbons, D.N. Page and C.N. Pope, {\it Einstein
metrics on $S^3$, $\R^3$ and $\R^4$ bundles}, Commun. Math. Phys.
{\bf 127}, 529 (1990).

\bm{ootyas} T. Oota and Y. Yasui,
{\it Explicit toric netric on resolved Calabi-Yau cone}, hep-th/0605129.

\bm{chlupo3} W. Chen, H. L\"u and C.N. Pope,
{\it General Kerr-NUT-AdS metrics in all dimensions}, hep-th/0604125.

\bm{gamaspwa} J.P. Gauntlett, D. Martelli, J. Sparks and D. Waldram,
{\it Sasaki-Einstein metrics on $S^2\times S^3$}, 
Adv. Theor. Math. Phys. {\bf 8}, 711 (2004), hep-th/0403002.

\bm{cvlupapo1} M. Cveti\v c, H. L\"u, D.N. Page and C.N. Pope,
{\it New Einstein-Sasaki spaces in five and higher dimensions},
Phys. Rev. Lett. {\bf 95}, 071101 (2005), hep-th/0504225.

\bm{cvlupapo2} M. Cveti\v c, H. L\"u, D.N. Page and C.N. Pope,
{\it New Einstein-Sasaki and Einstein spaces from Kerr-de Sitter},
hep-th/0505223.

\bm{chlupo1} W. Chen, H. L\"u and C.N. Pope,
{\it Kerr-de Sitter black holes with NUT charges}, hep-th/0601002.

\bm{lupova} H. L\"u, C.N. Pope and J.F. Vazquez-Poritz,
{\it A new construction of Einstein-Sasaki metrics in $D\ge7$},
hep-th/0512306.

\bm{gmsw2} J.P. Gauntlett, D. Martelli, J.F. Sparks and D. Waldram,
{\it A new infinite class of Sasaki-Einstein manifolds},
Adv. Theor. Math. Phys. {\bf 8}, 987 (2006), hep-th/0403038.

\bibitem{clpv} W. Chen, H. L\"u, C.N. Pope and J.F. V\'azquez-Poritz,
{\it A note on Einstein-Sasaki metrics in $D\ge 7$},
  Class. Quant. Grav.  {\bf 22}, 3421 (2005), hep-th/0411218.

\end{thebibliography}
\end{document}